\documentclass[twocolumn,showpacs,amsmath,amssymb]{revtex4}
\usepackage{graphicx}
\usepackage{amsfonts}
\usepackage{dcolumn}
\usepackage{bm}

\usepackage{CJK}

\begin{document}
\title{Optimal Allocation of Resources for Suppressing Epidemic Spreading on Networks}

\author{Hanshuang Chen$^{1}$} \email{chenhshf@ahu.edu.cn}

\author{Guofeng Li$^1$}

\author{Haifeng Zhang$^2$}

\author{Zhonghuai Hou$^{3}$}\email{hzhlj@ustc.edu.cn}

\affiliation{$^{1}$School of Physics and Materials Science, Anhui
University, Hefei, 230601, China \\$^2$School of Mathematical Science, Anhui University, Hefei, 230601, China \\
$^3$Hefei National Laboratory for Physical Sciences at Microscales
\& Department of Chemical Physics, University of
 Science and Technology of China, Hefei, 230026, China}

\date{\today}

\begin{abstract}
Efficient allocation of limited medical resources is crucial for
controlling epidemic spreading on networks. Based on the
susceptible-infected-susceptible model, we solve an optimization
problem as how best to allocate the limited resources so as to
minimize the prevalence, providing that the curing rate of each node
is positively correlated to its medical resource. By quenched
mean-field theory and heterogeneous mean-field (HMF) theory, we
prove that epidemic outbreak will be suppressed to the greatest
extent if the curing rate of each node is directly proportional to
its degree, under which the effective infection rate $\lambda$ has a
maximal threshold $\lambda_c^{opt}=1/\left\langle k \right\rangle$
where $\left\langle k \right\rangle$ is average degree of the
underlying network. For weak infection region
($\lambda\gtrsim\lambda_c^{opt}$), we combine a perturbation theory
with Lagrange multiplier method (LMM) to derive the analytical
expression of optimal allocation of the curing rates and the
corresponding minimized prevalence. For general infection region
($\lambda>\lambda_c^{opt}$), the high-dimensional optimization
problem is converted into numerically solving low-dimensional
nonlinear equations by the HMF theory and LMM. Counterintuitively,
in the strong infection region the low-degree nodes should be
allocated more medical resources than the high-degree nodes to
minimize the prevalence. Finally, we use simulated annealing to
validate the theoretical results.

\end{abstract}
 \maketitle

A challenging problem in epidemiology is how best to allocate
limited resources of treatment and vaccination so that they will be
most effective in suppressing or reducing outbreaks of epidemics.
This problem has been a subject of intense research in statistical
physics and many other disciplines
\cite{RevModPhys.87.925,IEEE.36.26}. Inspired by the percolation
theory, the simplest strategy is to randomly choose a fraction of
nodes to immunize. However, the random immunization is inefficient
for heterogeneous networks. Later on, many more effective
immunization strategies have been developed, ranging from global
strategies like targeted immunization based on node degree
\cite{PhysRevE.65.036104} or betweenness centrality
\cite{PhysRevE.65.056109} to local strategies, like acquaintance
immunization \cite{PhysRevLett.91.247901} and (bias) random walk
immunization \cite{EPL.68.908,PhysRevE.74.056105} and to some others
in between \cite{EPJB49.259}. Further improvements were done by
graph partitioning \cite{PhysRevLett.101.058701} and the
optimization of the susceptible size \cite{PhysRevE.84.061911}.
Besides the degree heterogeneity, community structure has also a
major impact on disease immunity
\cite{NJP11.123018,PlosCP6.e1000736}. Recently, a message-passing
approach was used to find an optimal set of nodes for immunization
\cite{PhysRevX.4.021024}. The immunization has been mapped onto the
optimal percolation problem \cite{Morone2015}. Based on the idea of
explosive percolation, an ``explosive immunization" method
 has been proposed \cite{PhysRevLett.117.208301}. However, some
diseases like the common cold and influenza that can be modeled by
the susceptible-infected-susceptible (SIS) model, do not confer
immunity and individuals can be infected over and over again. Under
the situations, one way to control the spread of the diseases is to
reduce the risk of the infection, such as adaptive rewiring links
incident to infected individuals \cite{PhysRevLett.96.208701} and
dynamical interplay between awareness and epidemic spreading
\cite{PhysRevLett.111.128701}.

An alternative way to control the epidemic spreading of SIS type by
designing an optimal strategy for distributing the limited medical
resources so as to suppress the epidemic outbreak to the greatest
extent and minimize the prevalence once the epidemic outbreak has
happened. It is reasonable to assume the curing rate of each node is
positively correlated to the medical resources allocated to it.
Therefore, the optimal allocation of medical resources is equivalent
to that of the curing rates. Assuming the total medical resources
are limited, the average curing rate is thus considered to be fixed.
This problem has been addressed as a constraint optimization problem
in several previous works. When the curing rate can be only tuned in
a fixed number of feasible values, this problem has been proved to
be NP-complete \cite{Prakash2013}. Instead, when the curing rate can
continuously varies in a given interval, some efficient algorithms
have been developed for minimizing the threshold of epidemic
outbreak \cite{Wan2008,Preciado2013} or the steady-state infection
density \cite{Gourdin2011}. In the present work, we theoretically
solve the constraint optimization problem in both epidemic-free and
endemic phases within the mean-field framework. On the one hand, we
prove that the epidemic outbreak can be suppressed to the most
extent when the curing rate of each node is directly proportional to
its degree, under which the epidemic threshold is maximized that is
the inverse of the average degree of the underlying network. On the
other hand, once the epidemic has broken out but close to the
threshold, we analytically show the optimal curing rate should be
adjusted in terms of the difference of node degree with average
degree and the distance to epidemic threshold. For the general
infection region, the optimization problem can be simplified to
solve three nonlinear equations.

To formulate our problem, we consider the SIS model on an undirected
network of size $N$. The network is described by an adjacency matrix
$\mathbb{A}$ whose entries are defined as $A_{ij}=1$ if nodes $i$
and $j$ are connected, and $A_{ij}=0$ otherwise. Each node is either
susceptible or infected. A susceptible node $i$ can be infected by
its infective neighbor with an infection rate $\beta$, and an
infected node $i$ recovers with a nonvanishing curing rate $\mu_i$.
Here, we consider that the curing rate is allowed to vary from one
node to another one. In general, the more available medical resource
of a node $i$ has, the larger $\mu_i$ is. Assuming that the total
amount of medicine resource is limited, the average curing rate is
thus fixed, i.e.,
\begin{eqnarray}
\left\langle {{\mu _i}} \right\rangle  = \mu {\kern 10pt}  and
{\kern 10pt} {\mu _i} \geqslant 0,{\kern 5pt} \forall i. \label{e1}
\end{eqnarray}
Our goal is to find out an optimal allocation of $\{\mu_i\}$ under
the constraint Eq.(\ref{e1}) so as to minimize the prevalence
$\rho$, that is the fraction of infected nodes.


In the quenched mean-field (QMF) theory, the probability $\rho_i(t)$
that node $i$ is infected at time $t$ is described by
$N$-intertwined equations \cite{Wang2003,Mieghem2009,Gomez2010},
\begin{eqnarray}
\frac{{d{\rho _i}(t)}}{{dt}} =  - {\mu _i}{\rho _i}(t) + \beta
\left[ {1 - {\rho _i}(t)} \right]\sum\limits_j {{A_{ij}}} {\rho
_j}(t). \label{e2}
\end{eqnarray}
In the steady state, $d\rho_i(t)/dt=0$, $\rho_i$ is determined by a
set of nonlinear equations,
\begin{eqnarray}
{\rho _i} = \frac{{\beta \sum\nolimits_j {{A_{ij}}{\rho _j}}
}}{{{\mu _i} + \beta \sum\nolimits_j {{A_{ij}}{\rho _j}}
}}.\label{e3}
\end{eqnarray}
One can notice that $\rho_i=0$ is always a solution of
Eq.(\ref{e2}). This trivial solution corresponds to an absorbing
state with no infective nodes. A nonzero solution $\rho_i>0$ exists
if the effective infection rate $\lambda=\beta/\mu$ is larger than
the so-called epidemic threshold $\lambda_c$. In this case, the
prevalence $\rho = \sum\nolimits_i {{\rho _i}} /N$ is nonzero
corresponding to an endemic state. By linear stability analysis for
Eq.(\ref{e2}) around $\rho_i=0$, $\lambda_c$ is determined by which
the largest eigenvalue of the matrix, $-\mathbb{U}+\beta
\mathbb{A}$, is zero, where $\mathbb{U}= diag(\mu_i)$ is a diagonal
matrix. For the standard SIS model, $\mu_i\equiv\mu$ for all $i$,
one can immediately obtain the well-known result,
$\lambda_{c,QMF}^{sta}=1/\Lambda_{max}(\mathbb{A})$ with the largest
eigenvalue of the adjacency matrix $\Lambda_{max}(\mathbb{A})$. In
our SIS model, the outbreak of epidemics will be suppressed to the
greatest extent, which implies that the epidemic threshold of the
optimal SIS model will be maximized.


For this purpose, we first decompose the diagonal matrix
$\mathbb{U}$ into two diagonal matrices, $\mathbb{U}=\mathbb{\bar
U}+ \Delta \mathbb{U}$, where $ \mathbb{\bar U}= diag\{\mu
k_i/\left\langle k \right\rangle \}$ with $k_i$ being the degree of
node $i$ and $\Delta \mathbb{U}= diag\{\Delta \mu_i\}$. Since
$Tr(\mathbb{U})=Tr(\mathbb{\bar U})=N\mu$, $\Delta \mathbb{U}$ must
satisfy the constraint $Tr(\Delta \mathbb{U})=0$. For the real
symmetric matrix, $\mathbb{U}-\beta \mathbb{A}$, its largest
eigenvalue $\Lambda_{max}$ satisfies the following inequality,
\begin{eqnarray}
\Lambda_{max}\geq \textbf{v}^T (-\mathbb{U}+\beta \mathbb{A})
\textbf{v},\label{e4}
\end{eqnarray}
where $\textbf{v}$ is a column vector satisfying
$\textbf{v}\in\mathbb{R}^N$ and $||\textbf{v}||=1$. If we set
$\textbf{v}=\frac{1}{{\sqrt N }}{\left( {1, \cdots ,1} \right)^T}$,
Eq.(\ref{e4}) becomes
\begin{eqnarray}
\Lambda_{max}\geq \textbf{v}^T (-\mathbb{\bar U}+\beta
\mathbb{A})\textbf{v}- \textbf{v}^T \Delta \mathbb{U} \textbf{v}
=-\mu+\beta \left\langle k \right\rangle.\label{e5}
\end{eqnarray}
Since $\Lambda_{max}=0$ at the epidemic threshold, Eq.(\ref{e5})
leads to an upper bound of epidemic threshold,
$\lambda_c\leq1/\left\langle k \right\rangle$. The condition that
the epidemic threshold equals to the upper bound holds when
$\textbf{v}$ is the eigenvector of $\mathbb{U}-\beta \mathbb{A}$
corresponding to its largest eigenvalue. If we set
$\mathbb{U}=\mathbb{\bar U}$ and $\beta=\mu/\left\langle k
\right\rangle$, $-\mathbb{U}+\beta \mathbb{A}=-\mu/\left\langle k
\right\rangle \mathbb{L}$, where $\mathbb{L}$ is the Laplacian
matrix of the underlying network. It is well-known that the smallest
eigenvalue of $\mathbb{L}$ is zero and the corresponding eigenvector
is $\textbf{v}$. Therefore, if the curing rate of each node is
directly proportional to its degree, i.e.,
\begin{eqnarray}
{\mu _i} = \mu_i^*=\mu \frac{{ {k_i}}}{\left\langle k
\right\rangle},\label{e6}
\end{eqnarray}
the epidemic threshold will be maximized,
\begin{eqnarray}
\lambda_{c,QMF}^{opt}=\frac{1}{\left\langle k
\right\rangle}.\label{e7}
\end{eqnarray}
In the QMF theory, the epidemic threshold of the optimal SIS model
is no less than that of the standard SIS model,
$\lambda_{c,QMF}^{opt} \geq \lambda_{c,QMF}^{sta}$, as the lower
bound of $\Lambda_{max}(\mathbb{A})$ is $\left\langle k
\right\rangle$ for any types of networks \cite{Mieghem_Book2011}.

The above results can be also derived from the heterogeneous
mean-field (HMF) theory. In the framework of HMF, these nodes with
the same degree are considered to be statistically equivalent. The
constraint Eq.(\ref{e1}) becomes
\begin{eqnarray}
\left\langle {{\mu _k}} \right\rangle  = \sum\nolimits_k {P(k)} {\mu
_k}{\kern 1pt}  = \mu {\kern 1pt} {\kern 1pt} {\kern 1pt} {\kern
1pt} {\kern 1pt} {\kern 1pt} {\kern 1pt} {\kern 1pt} {\kern 1pt}
{\kern 1pt} {\kern 1pt} and{\kern 1pt} {\kern 1pt} {\kern 1pt}
{\kern 1pt} {\kern 1pt} {\kern 1pt} {\kern 1pt} {\kern 1pt} {\kern
1pt} {\kern 1pt} {\mu _k} \geqslant 0,{\kern 1pt} {\kern 1pt} {\kern
1pt} {\kern 1pt} \forall k{\kern 1pt},\label{e8}
\end{eqnarray}
where $\mu_k$ is the curing rate of nodes of degree $k$, and $P(k)$
is the degree distribution. Some related works have studied the SIS
model \cite{PhysRevE.65.055103} and its metapopulation version
\cite{PhysRevE.86.036114} with the curing rate, $\mu_k\sim
k^\alpha$, but such a power-law form did not guarantee to be the
optimal one. In \cite{Borgs2010}, the authors consider a simple
heuristic strategy to control epidemic extinction where the curing
rate is directly proportional to node degree. They showed that on
any graph with bounded degree the extinction time is sublinear with
the size of the network. Further improvement has been done by a
heuristic PageRank algorithm to allocate curing rates based on the
initial condition of infected nodes \cite{Chung2009}. The present
study does not require any assumptions about the form of the cure
rate with node degree in advance except to the constraint
Eq.(\ref{e8}). The dynamical evolution of $\rho_k(t)$, the
probability of nodes of degree $k$ being infected at time $t$, reads
\cite{PRL01003200},
\begin{eqnarray}
\frac{{d{\rho _k}\left( t \right)}}{{dt}} =  - {\mu _k}{\rho
_k}\left( t \right) + \beta \left[ {1 - {\rho _k}\left( t \right)}
\right]k\Theta(t),\label{e9}
\end{eqnarray}
where $\Theta$ is the probability of finding an infected node
following a randomly chosen edge. In the case of uncorrelated
networks, $\Theta(t)$ can be written as
\begin{eqnarray}
\Theta(t)  = \sum\limits_k {\frac{{kP(k)}}{{\left\langle k
\right\rangle }}} {\rho _k(t)}.\label{e10}
\end{eqnarray}
In the steady state, $d\rho_k(t)/dt=0$, Eq.(\ref{e9}) becomes
\begin{eqnarray}
{\rho _k} = \frac{{\beta k\Theta }}{{{\mu _k} +  \beta k\Theta
}}.\label{e11}
\end{eqnarray}
Substituting Eq.(\ref{e11}) into Eq.(\ref{e10}), we obtain a
self-consistent equation of $\Theta$,
\begin{eqnarray}
\Theta  = \sum\limits_k {\frac{{kP(k)}}{{\left\langle k
\right\rangle }}\frac{{\beta k\Theta }}{{{\mu _k} +  \beta k\Theta
}}}.\label{e12}
\end{eqnarray}


The epidemic threshold is determined by which the derivation of the
r.h.s of Eq.(\ref{e12}) with respect to $\Theta$ at $\Theta=0$
equals to one, leading to
\begin{eqnarray}
{\beta _{c,HMF}} = \frac{{\left\langle k \right\rangle
}}{{\sum\nolimits_k {\frac{{{k^2}P(k)}}{{{\mu _k}}}} }}.\label{e13}
\end{eqnarray}
For a given $P(k)$, maximizing $\beta_c$ is equivalent to minimizing
the denominator of the r.h.s of Eq.(\ref{e13}). For this purpose, we
employ Lagrange multiplier method (LMM) to maximize the epidemic
threshold, where the Lagrange function is written as,
\begin{eqnarray}
\mathcal {L} = \sum\limits_k {\frac{{{k^2}P(k)}}{{{\mu _k}}}}  +
\tau \left( {\sum\limits_k {P(k)} {\mu _k} - \mu }
\right),\label{e14}
\end{eqnarray}
where $\tau$ is called the Lagrange multiplier. Taking the
derivation of $\mathcal {L}$ with respect to $\mu_k$,
\begin{eqnarray}
\frac{{\partial \mathcal {L}}}{{\partial {\mu _k}}} =  -
\frac{{{k^2}P(k)}}{{\mu _k^2}} + \tau P(k),\label{e15}
\end{eqnarray}
and letting ${{\partial \mathcal {L}} \mathord{\left/
 {\vphantom {{\partial L} {\partial {\mu _k} = 0}}} \right.
 \kern-\nulldelimiterspace} {\partial {\mu _k} = 0}}$ combined with Eq.(8), we arrive at a maximal epidemic threshold
\begin{eqnarray}
\lambda _{c,HMF}^{opt} = \frac{1}{{\left\langle k \right\rangle
}},\label{e16}
\end{eqnarray}
and the corresponding allocation of $\{\mu_k\}$,
\begin{eqnarray}
{\mu _k} = \mu_k^*=\mu \frac{k}{{\left\langle k \right\rangle
}}\label{e17}
\end{eqnarray}
Interestingly, the HMF results are consistent with the QMF ones.
Also, in the HMF theory the epidemic threshold of the optimal SIS
model is no less than that of the standard SIS model,
$\lambda_{c,HMF}^{opt} \geq \lambda_{c,HMF}^{sta}=\left\langle k
\right\rangle/\left\langle k^2 \right\rangle$.

For $\lambda$ larger than but close to $\lambda_c^{opt}$,
$\lambda\gtrsim\lambda_c^{opt}$, we shall combine a perturbation
theory with LMM to optimize the prevalence. To the end, we assume
that for $\lambda = \lambda_c^{opt}+ \Delta \lambda$,
$\mu_k=\mu_k^*+\Delta \mu_k$ and $\Theta=\Theta^*+\Delta \Theta$,
where $\Theta^*=1 - \frac{\mu }{{\beta \left\langle k \right\rangle
}}$ is the solution of Eq.(\ref{e12}) for $\mu_k=\mu_k^*$. Expanding
Eq.(\ref{e12}) around ${\left( {\mu _k^*,{\Theta ^*}} \right)}$ to
the second order, and then using the constraint $\sum\nolimits_k
P(k)\Delta \mu_k=0$ and simultaneously ignoring the second-order
small quantity $\Delta \Theta^2$, it yields [See Appendix A for
details]
\begin{eqnarray}
\Delta \Theta  = \frac{1}{{{\beta ^2}\left\langle k \right\rangle
}}\sum\limits_k {\frac{{P(k)}}{k}} \Delta \mu _k^2{\kern
1pt}.\label{e18}
\end{eqnarray}
Around ${\left( {\mu _k^*,{\Theta ^*}} \right)}$, the change $\Delta
\rho$ in the prevalence $\rho=\sum\nolimits_k P(k)\rho _k$ can be
written as
\begin{eqnarray}
\Delta \rho  =  - \frac{{{\Theta ^*}}}{\beta }\sum\limits_k
{\frac{{P(k)}}{k}} \Delta {\mu _k} + \left( {1 - {\Theta ^*}}
\right)\Delta \Theta.\label{e19}
\end{eqnarray}
Again using LMM to minimize $\Delta \rho$ under the constraints
$\sum\nolimits_k P(k)\Delta \mu_k=0$ and Eq.(\ref{e18}), we obtain a
minimal $\rho=\rho^*+\Delta \rho^{opt}$ with ${\rho ^*}
=\sum\nolimits_k {P(k)\frac{{\beta k{\Theta ^*}}}{{\mu _k^* + \beta
k{\Theta ^*}}}}$ and
\begin{eqnarray}
\Delta \rho^{opt}  &=     - \frac{1}{{4\lambda }}{\left\langle k
\right\rangle ^2}\left( {\left\langle {{k^{ - 1}}} \right\rangle -
{{\left\langle k \right\rangle }^{ - 1}}} \right)\Delta {\lambda ^2}
\nonumber \\
 &\simeq  - \frac{1}{4}{\left\langle k \right\rangle
^3}\left( {\left\langle {{k^{ - 1}}} \right\rangle  - {{\left\langle
k \right\rangle }^{ - 1}}} \right)\Delta {\lambda ^2}.\label{e20}
\end{eqnarray}
Since $\left\langle {{k^{ - 1}}} \right\rangle  > {\left\langle k
\right\rangle ^{ - 1}}$ for any degree inhomogeneous networks in
terms of Jensen's inequality, $\Delta \rho^{opt}<0$ and thus $\rho$
will be reduced. The corresponding optimal allocation
$\mu_k=\mu_k^*+\Delta \mu_k$ with
\begin{eqnarray}
\Delta {\mu _k} = \frac{\mu }{2}\left\langle k \right\rangle \lambda
 \left( {\left\langle k \right\rangle  - k} \right) \Delta \lambda
\simeq  \frac{\mu }{2} \left( {\left\langle k \right\rangle-k}
\right)\Delta \lambda.\label{e21}
\end{eqnarray}
This implies that as $\lambda$ is increased from $\lambda_c^{opt}$,
the curing rates of the nodes with degrees less than the average
degree will be increased, while the curing rates of the nodes with
degrees larger than the average degree will be decreased. The
amplitude of the change will depend on the difference between the
degree of each node and the average degree, $\left\langle k
\right\rangle-k$, and the distance of the effective infection rate
to its critical value, $\Delta \lambda$.

For $\lambda$ is larger than but not close to $\lambda_c^{opt}$,
$\lambda>\lambda_c^{opt}$, since the nonlinear characteristic of the
model, analytical expression of optimal allocation of $\{\mu_k\}$
and the corresponding the minimal $\rho$ is almost impossible.
However, with the aid of HMF theory and LMM, the high-dimensional
optimization problem can be converted to numerically solving the
low-dimensional nonlinear equations [See Appendix B for details]. In
the general infection region, $\mu_k$ satisfies the following
equation,
\begin{eqnarray}
\mu _k = \left\{ \begin{gathered}
  \sqrt {\frac{{\beta k\Theta }}{\tau } + \frac{{\kappa \beta {k^2}}}{{\tau \left\langle k \right\rangle }}}  - \beta k\Theta >0,  \; k <k_c \hfill \\
  0, \; k \geq k_c \hfill \\
\end{gathered}  \right.\label{e22}
\end{eqnarray}
where $\tau$ and $\kappa$ are the Lagrange multipliers, and $k_c$ is
a threshold degree to guarantee $\mu_k>0$ for $k<k_c$ and it will be
determined later. $\Theta$, $\tau$ and $\kappa$ are determined by
the following three equations,
\begin{eqnarray}
\sqrt {\frac{{\beta \tau \left\langle k \right\rangle }}{\Theta }}
\sum\limits_{k = {k_{\min }}}^{{k_{\max }}} {\sqrt \xi  P(k)}  -
\beta \tau \left\langle k \right\rangle \sum\limits_{k = {k_{\min
}}}^{{k_{\max }}} {\xi P(k)}  \nonumber \\ - \beta \kappa \tau
\sum\limits_{k = {k_{\min }}}^{{k_{\max }}} {k\xi P(k)}   -
\frac{\kappa }{{\left\langle k \right\rangle \Theta }}\sum\limits_{k
= {k_c}}^{{k_{\max }}} {kP(k) = 0},\label{e23}
\end{eqnarray}
\begin{eqnarray}
\mu  =  \sqrt {\frac{{\beta \Theta }}{{\tau \left\langle k
\right\rangle }}} \sum\limits_{k = {k_{\min }}}^{{k_c}} {k
\xi^{-\frac{1}{2}}P(k)} -\beta \Theta \sum\limits_{k =
{k_{min}}}^{{k_{c}}}{k P(k)},\label{e24}
\end{eqnarray}
\begin{eqnarray}
\Theta= \sqrt {\frac{{\beta \tau \Theta }}{{\left\langle k
\right\rangle }}} \sum\limits_{k = {k_{\min }}}^{{k_c}} {k^2
\xi^{-\frac{1}{2}}P(k)} + \frac{1}{{\left\langle k \right\rangle
}}\sum\limits_{k = {k_c}}^{{k_{\max }}} {k P(k)},\label{e25}
\end{eqnarray}
where we have used $\xi=k/({\left\langle k\right\rangle+\kappa k})$.

To numerically solve $\Theta$, $\tau$ and $\kappa$ by
Eqs.(\ref{e23},\ref{e24},\ref{e25}), $k_c$ is needed to be known in
advance. To the end, we adopt a numerical scheme as follows. (i)
Firstly we set $k_c=k_{max}$ where $k_{max}$ is the maximal degree
of the underlying network; (ii) we numerically solve $\Theta$,
$\tau$ and $\kappa$ by Eqs.(\ref{e23},\ref{e24},\ref{e25}), and then
test the condition $\mu_k>0$ for all $k<k_c$ by Eq.(\ref{e22});
(iii) if the condition is not satisfied, $k_c$ will be decreased by
$k_c\leftarrow k_c-1$ and return to ii) until the condition
Eq.(\ref{e22}) is fulfilled.

\begin{figure}
\centerline{\includegraphics*[width=1.0\columnwidth]{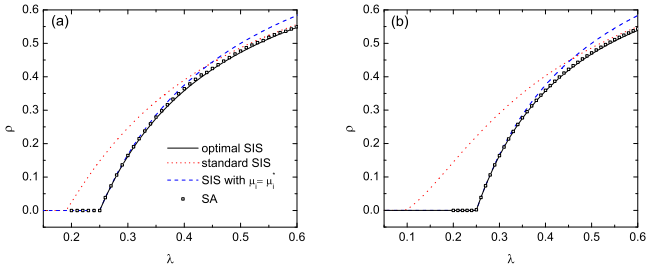}}
\caption{(color online). Prevalence $\rho$ versus the effective
infection rate $\lambda$ in ER networks (a) and BA networks (b) with
equal $N=1000$ and $\left\langle k \right\rangle=4$. The solid lines
correspond to the results from the optimal SIS model, the dotted
line to the results of the standard SIS model, and the dashed line
to the SIS model with $\mu_i=\mu_i^*$. The squares correspond to the
results from SA. \label{fig1}}
\end{figure}



Figure \ref{fig1} shows the optimized results of $\rho$ as a
function $\lambda$ (solid line) in Erd\"os-R\'enyi (ER) random
networks (a) and Barab\'asi-Albert scale-free networks (b) with
equal network size $N=1000$ and average degree $\left\langle k
\right\rangle=4$. For comparison, we also show the results of the
standard SIS model (dotted line) and of the SIS model with the
curing rates $\mu_i=\mu_i^*$ (dashed line). As expected by the
theoretical prediction, the epidemic threshold of the optimal SIS
model $\lambda_c^{opt}=1/\left\langle k \right\rangle$, which is
significantly larger than that of the standard SIS model, but
coincides with the case of $\mu_i=\mu_i^*$. While for
$\lambda>\lambda_c^{opt}$, the prevalence for $\mu_i=\mu_i^*$ is
always larger than the optimal choice, and even larger than the
standard SIS model in the strong infection region, indicating that
$\mu_i=\mu_i^*$ is not a good choice once the epidemic outbreak has
happened.

We use the simulated annealing (SA) technique to validate our
theoretical results. The SA builds a Monte Carlo Markov Chain that
in the long run converges to the minimum of a given energy function
$\mathcal {E}$, where $\mathcal {E}=\rho$ can be obtained by
numerically iterating Eq.(\ref{e3}). The main steps of SA are as
follows. At beginning, we assign to a given set of $\{\mu_i\}$
satisfying the constraint Eq.(\ref{e1}) (e.g., $\mu_i=\mu$ for all
$i$). Then, we randomly choose two distinct nodes, say $i$ and $j$,
and try to make the changes $\mu_i \leftarrow \mu_i +\delta$ and
$\mu_j \leftarrow \mu_j -\delta$ with the standard Metropolis
probability $\min(1, e^{-\beta_{SA}\Delta\mathcal {E}})$, where
$\delta$ is randomly chosen between $-\mu_i$ and $\mu_i+\mu_j$ to
guarantee the curing rate is always not less than zero. $\beta_{SA}$
is the inverse temperature of SA which slowly increases from
$10^{-2}$ to $10^{4}$ via an annealing protocol. $\Delta \mathcal
{E}$ is the change of the energy function $\mathcal {E}$ due the
change of $\mu_i$ and $\mu_j$, We tested several different annealing
protocols and we adopted one in which the inverse temperature of SA
$\beta_{SA}$ is updated by $\beta_{SA} \leftarrow 1.01 \beta_{SA}$
after each $N$ attempts for updating $\{\mu_i\}$. The SA results are
also shown in Fig.1 (square dots), which agree with the theoretical
prediction.

\begin{figure}
\centerline{\includegraphics*[width=1.0\columnwidth]{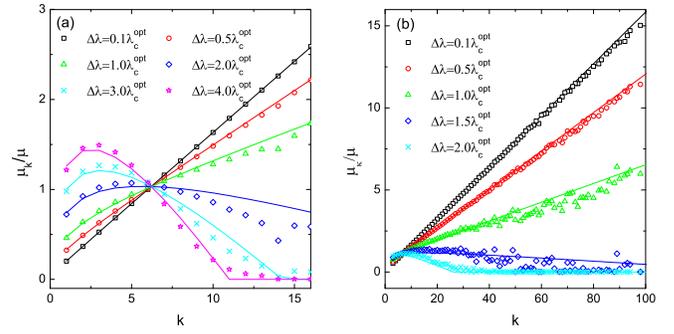}}
\caption{(color online). The optimal allocation of $\{\mu_k\}$ as a
function of node degree $k$ for several distinct $\lambda$ in ER
random networks (a) and BA scale-free networks (b) with equal
$N=1000$ and $\left\langle k \right\rangle=6$. The lines and dots
indicate the theoretical and SA results, respectively. \label{fig2}}
\end{figure}

In Fig.\ref{fig2}, we show the optimal allocation of $\{\mu_k\}$ as
a function of node degree $k$ for several distinct $\lambda$ in ER
random networks (a) and BA scale-free networks (b), in which the
theoretical results and the SA ones are indicated by the lines and
dots, respectively. For $\lambda \gtrsim \lambda_c^{opt}$, $\mu_k$
increases linearly as $k$ with the slope depending on the distance
to the epidemic threshold. The results have been well predicted by
Eq.(\ref{e17}) and Eq.(\ref{e21}). For the region away from the
threshold, $\mu_k$ will deviate from linear relation with $k$. For
sufficiently large $\lambda$, $\mu_k$ for large $k$ can be less than
that for small $k$, and even $\mu_k$ vanishes when $k$ exceeds a
threshold value, as given by Eq.(\ref{e22}). This surprising result
implies that in the strong infection region more medicine resources
should be put into these low-degree nodes other than high-degree
nodes.

In conclusion, we have theoretically studied a constraint
optimization problem as how best to distribute the limited medicine
resources (curing rates) for controlling the epidemics of SIS type.
Based on the QMF and HMF theories, we have shown that the optimal
allocation lies in the effective infection rate $\lambda$ (or the
basic reproduction number $R_0=\left\langle k \right\rangle
\lambda$). If $R_0\leqslant 1$, the curing rate of each node should
be in direct proportion to its degree, under which the epidemic
outbreak will be suppressed to the most extent and the epidemic
threshold will be maximized, Eq.(\ref{e7}) or Eq.(\ref{e16}). Once
the maximal epidemic threshold is just across ($R_0\gtrsim1$), the
epidemic will spread persistently. In this case, we have
analytically shown that the change in the curing rate of each node
depends linearly on the difference between the average degree and
its degree and the distance to epidemic threshold, Eq.(\ref{e21}).
For the general infection region ($R_0>1$), it is almost impossible
to derive an analytical solution of the optimization problem;
however, it can be simplified to an much more easily problem of
numerical calculation of three nonlinear equations,
Eqs.(\ref{e23},\ref{e24},\ref{e25}). Surprisingly, we found that in
the strong infection region the curing rates of the low-degree nodes
can overpower those of the high-degree nodes to ensure the
minimization of the prevalence.

An interesting generalization is how to solve the present constraint
optimization problem based on other existing theoretical methods,
such as pair mean-field method that takes into account the role of
dynamical correlations between neighboring nodes
\cite{Eames2002,PhysRevLett.107.068701,PhysRevLett.111.068701,Mata2014,PhysRevLett.115.078701,PhysRevLett.116.258301,PhysRevE.85.056111,Mata2103,Kiss2015}.
Moreover, the method presented here could be applied to a number of
other optimization problems, for example, controlling opinion
dynamics in social networks \cite{RMP09000591}. This will be the
subject of future work.



\emph{Acknowledgments}: This work was supported by National Science
Foundation of China (Grants Nos. 11205002, 61473001, 21673212), the
Key Scientific Research Fund of Anhui Provincial Education
Department (Grant No. KJ2016A015) and ``211" Project of Anhui
University (Grant No. J01005106).

\bibliographystyle{apsrev}

\appendix
\section{Weak infection region}
For $\lambda$ larger than but close to $\lambda_c^{opt}$, $\lambda
\gtrsim \lambda_c^{opt}$, we have combined a perturbation theory
with Lagrange multiplier method (LMM) to optimize the prevalence
$\rho$. For $\lambda = \lambda_c^{opt}+ \Delta \lambda$, we have
$\mu_k=\mu_k^*+\Delta \mu_k$ and $\Theta=\Theta^*+\Delta \Theta$,
where $\mu_k^*=\mu k/\left\langle k \right\rangle$, and $\Theta^*=1
- \frac{\mu }{{\beta \left\langle k \right\rangle }}$ is the
solution of self-consistent equation of $\Theta$, Eq.(\ref{e12}) in
the main text, under $\mu_k=\mu_k^*$. Since $\Theta>0$ in the region
of epidemic spreading, Eq.(\ref{e12}) in the main text can be
rewritten as
\begin{equation}\label{eq1}\tag{S1}
\frac{\beta }{{\left\langle k \right\rangle }}\sum\limits_k
{\frac{{{k^2}P(k)}}{{{\mu _k} + \beta k\Theta }} }=1.
\end{equation}
Expanding the above equation around ${\left( {\mu _k^*,{\Theta ^*}}
\right)}$ to the second-order, it yields
\begin{equation}\label{eq2}\tag{S2}
\begin{split}
&\sum\limits_k {{{\left. {\frac{{\partial f}}{{\partial {\mu _k}}}}
\right|}_{\left( {\mu _k^*,{\Theta ^*}} \right)}}} \Delta {\mu _k} +
{\left. {\frac{{\partial f}}{{\partial \Theta }}} \right|_{\left(
{\mu _k^*,{\Theta ^*}} \right)}}\Delta \Theta  \\&+
\frac{1}{2}\sum\limits_k {{{\left. {\sum\limits_{k'}
{\frac{{{\partial ^2}f}}{{\partial {\mu _k}\partial {\mu _{k'}}}}} }
\right|}_{\left( {\mu _k^*,{\Theta ^*}} \right)}}} \Delta {\mu
_k}\Delta {\mu _{k'}}{\kern 1pt} {\kern 1pt} {\kern 1pt} \nonumber
\\ & + \sum\limits_k {{{\left. {\frac{{{\partial
^2}f}}{{\partial {\mu _k}\partial \Theta }}} \right|}_{\left( {\mu
_k^*,{\Theta ^*}} \right)}}} \Delta {\mu _k}\Delta \Theta  + {\left.
{\frac{1}{2}\frac{{{\partial ^2}f}}{{\partial \Theta \partial \Theta
}}} \right|_{\left( {\mu _k^*,{\Theta ^*}} \right)}}\Delta \Theta^2
=0,
\end{split}
\end{equation}
where $f \buildrel \Delta \over = \frac{\beta }{{\left\langle k
\right\rangle }}\sum\limits_k {\frac{{{k^2}P(k)}}{{{\mu _k} + \beta
k\Theta }} - } 1$, and
\begin{equation}\label{eq3}\tag{S3}
\begin{split}
&{\left. {\frac{{\partial f}}{{\partial {\mu _k}}}} \right|_{\left( {\mu _k^*,{\Theta ^*}} \right)}} =   - \frac{P(k)}{{\beta \left\langle k \right\rangle }}\\
&{\left. {\frac{{\partial f}}{{\partial \Theta }}} \right|_{\left( {\mu _k^*,{\Theta ^*}} \right)}} =   - 1\\
&{\left. {\frac{{{\partial ^2}f}}{{\partial {\mu _k}\partial {\mu _{k'}}}}} \right|_{\left( {\mu _k^*,{\Theta ^*}} \right)}} = {\delta _{kk'}}\frac{{2P(k)}}{{{\beta ^2}\left\langle k \right\rangle k}}\\
&{\left. {\frac{{{\partial ^2}f}}{{\partial {\mu _k}\partial \Theta }}} \right|_{\left( {\mu _k^*,{\Theta ^*}} \right)}} = \frac{{2P(k)}}{{\beta \left\langle k \right\rangle }}\\
&{\left. {\frac{{{\partial ^2}f}}{{\partial \Theta \partial \Theta
}}} \right|_{\left( {\mu _k^*,{\Theta ^*}} \right)}} = 2.
\end{split}
\end{equation}
Substituting Eq.(\ref{eq3}) into Eq.(\ref{eq2}), we obtain
\begin{equation}\label{eq4}\tag{S4}
\begin{split}
& - \frac{1}{{\beta \left\langle k \right\rangle }}\sum\limits_k
{P(k)} \Delta {\mu _k} - \Delta \Theta  + \frac{1}{{{\beta
^2}\left\langle k \right\rangle }}\sum\limits_k {\frac{{P(k)}}{k}}
\Delta \mu _k^2{\kern 1pt}
\\&+ \frac{2}{{\beta \left\langle k \right\rangle }}\sum\limits_k
{P(k)} \Delta {\mu _k}\Delta \Theta  + \Delta {\Theta ^2} = 0.
\end{split}
\end{equation}
Using the constraint $\sum\nolimits_k P(k)\Delta \mu_k=0$ and
ignoring the second-order small quantity $\Delta \Theta^2\ll\Delta
\Theta$, Eq.(\ref{eq4}) becomes
\begin{equation}\label{eq5}\tag{S5}
\Delta \Theta  = \frac{1}{{{\beta ^2}\left\langle k \right\rangle
}}\sum\limits_k {\frac{{P(k)}}{k}} \Delta \mu _k^2{\kern 1pt}.
\end{equation}
Around ${\left( {\mu _k^*,{\Theta ^*}} \right)}$, the change $\Delta
\rho$ in the prevalence $\rho=\sum\nolimits_k P(k)\rho _k$ can be
expanded in the leading order
\begin{equation}\label{eq6}\tag{S6}
\Delta \rho  = \sum\limits_k {{{\left. {\frac{{\partial \rho
}}{{\partial {\mu _k}}}} \right|}_{\left( {\mu _k^*,{\Theta ^*}}
\right)}}} \Delta {\mu _k} + {\left. {\frac{{\partial \rho
}}{{\partial \Theta }}} \right|_{\left( {\mu _k^*,{\Theta ^*}}
\right)}}\Delta \Theta,
\end{equation}
where
\begin{equation}\label{eq7}\tag{S7}
\begin{split}
  &{\left. {\frac{{\partial \rho }}{{\partial {\mu _k}}}} \right|_{\left( {\mu _k^*,{\Theta ^*}} \right)}} =  - \frac{{P(k){\Theta ^*}}}{{\beta k}} \hfill, \\
  &{\left. {\frac{{\partial \rho }}{{\partial \Theta }}} \right|_{\left( {\mu _k^*,{\Theta ^*}} \right)}} = 1 - {\Theta ^*} \hfill. \\
\end{split}
\end{equation}
Substituting Eq.(\ref{eq7}) into Eq(\ref{eq6}), we obtain
\begin{equation}\label{eq8}\tag{S8}
\Delta \rho  =  - \frac{{{\Theta ^*}}}{\beta }\sum\limits_k
{\frac{{P(k)}}{k}} \Delta {\mu _k} + \left( {1 - {\Theta ^*}}
\right)\Delta \Theta.
\end{equation}
In the following we use LMM to minimize $\Delta \rho$ under the
constraints $\sum\nolimits_k P(k)\Delta \mu_k=0$ and Eq.(\ref{eq5}).
Note that the first constraint is due to the fixed average curing
rate, and the second one is the requirement of the HMF dynamics.
Utilizing Eq.(\ref{eq8}) and the two constraints, the Lagrange
function can be written as
\begin{equation}\label{eq9}\tag{S9}
\begin{split}
\mathcal {L} =  &- \frac{{{\Theta ^*}}}{\beta }\sum\limits_k
{\frac{{P(k)}}{k}} \Delta {\mu _k} + \left( {1 - {\Theta ^*}}
\right)\Delta \Theta  \\&+ \tau \left( { - \Delta \Theta  +
\frac{1}{{{\beta ^2}\left\langle k \right\rangle }}\sum\limits_k
{\frac{{P(k)}}{k}} \Delta \mu _k^2} \right){\kern 1pt}  + \kappa
\sum\limits_k {P(k)} \Delta {\mu _k},
\end{split}
\end{equation}
where $\tau$ and $\kappa$ are the Lagrange multipliers. Taking the
derivative of $\mathcal {L}$ with respect to $\Delta \Theta$ and
$\Delta \mu_k$, we obtain
\begin{equation}\label{eq10}\tag{S10}
\frac{{\partial \mathcal {L}}}{{\partial \Delta \Theta }} = \left(
{1 - {\Theta ^*}} \right) - \tau,
\end{equation}
and
\begin{equation}\label{eq11}\tag{S11}
\frac{{\partial \mathcal {L}}}{{\partial \Delta {\mu _k}}} =  -
\frac{{{\Theta ^*}}}{\beta }\frac{{P(k)}}{k} + \tau \frac{1}{{{\beta
^2}\left\langle k \right\rangle }}\frac{{2P(k)}}{k}\Delta {\mu _k} +
\kappa P(k).
\end{equation}
Letting $\frac{{\partial L}}{{\partial \Delta \Theta }} =0$ and
$\frac{{\partial L}}{{\partial \Delta \mu_k }} =0$, we obtain
\begin{equation}\label{eq12}\tag{S12}
\tau  = 1 - {\Theta ^*},
\end{equation}
and
\begin{equation}\label{eq13}\tag{S13}
 - \frac{{{\Theta ^*}}}{{\beta k}} + \frac{{2\tau }}{{{\beta
^2}k\left\langle k \right\rangle }}\Delta {\mu _k} + \kappa  = 0,
\end{equation}
respectively. Substituting Eq.(\ref{eq13}) into the constraint
$\sum\nolimits_k P(k)\Delta \mu_k=0$, we obtain
\begin{equation}\label{eq14}\tag{S14}
\kappa  = \frac{{{\Theta ^*}}}{{\beta \left\langle k \right\rangle
}}.
\end{equation}
Combining Eqs.(\ref{eq12},\ref{eq13},\ref{eq14}), we obtain
\begin{equation}\label{eq15}\tag{S15}
\Delta {\mu _k} = \frac{\mu }{2}\left\langle k \right\rangle \lambda
 \left( {\left\langle k \right\rangle  - k} \right) \Delta \lambda
\simeq  \frac{\mu }{2} \left( {\left\langle k \right\rangle  - k}
\right)\Delta \lambda.
\end{equation}
Substituting Eq.(\ref{eq5}) and Eq.(\ref{eq15}) into Eq.(\ref{eq8}),
we obtain
\begin{equation}\label{eq16}\tag{S16}
\begin{split}
\Delta \rho^{opt}  = - \frac{1}{{4\lambda }}{\left\langle k
\right\rangle ^2}\left( {\left\langle {{k^{ - 1}}} \right\rangle -
{{\left\langle k \right\rangle }^{ - 1}}} \right)\Delta {\lambda ^2}
 \\ \simeq  - \frac{1}{4}{\left\langle k \right\rangle
^3}\left( {\left\langle {{k^{ - 1}}} \right\rangle  - {{\left\langle
k \right\rangle }^{ - 1}}} \right)\Delta {\lambda ^2}.
\end{split}
\end{equation}

\section{General Infection Region}
For $\lambda$ is larger than but not close to $\lambda_c^{opt}$,
since the nonlinear character of the model, analytical expression of
optimal allocation of $\{\mu_k\}$ and the corresponding the minimal
$\rho$ is in general impossible. However, with the aid of HMF theory
and LMM, the high-dimensional optimization problem can be converted
to numerically solving low-dimensional nonlinear equations. We first
write a Lagrange function as
\begin{equation}\label{eq17}\tag{S17}
\begin{split}
\mathcal {L}=  & \sum\limits_k {P(k)} \frac{{\beta k\Theta }}{{{\mu
_k} + \beta k\Theta }} + \tau \left( {\sum\limits_k {P(k)} {\mu _k}
- \mu } \right) \\ &+ \kappa \left( {\sum\limits_k
{\frac{{kP(k)}}{{\left\langle k \right\rangle }}} \frac{{\beta k
\Theta }}{{{\mu _k} + \beta k\Theta }} - \Theta } \right),
\end{split}
\end{equation}
where $\tau$ and $\kappa$ are the Lagrange multipliers. Taking the
derivative of $\mathcal {L}$ with respect to $\mu_k$ and $\Theta$,
we obtain
\begin{equation}\label{eq18}\tag{S18}
\frac{{\partial \mathcal {L}}}{{\partial {\mu _k}}} =  -
P(k)\frac{{\beta k\Theta }}{{{{\left( {{\mu _k} + \beta k\Theta }
\right)}^2}}} + \tau P(k)- \kappa \frac{{kP(k)}}{{\left\langle k
\right\rangle }}\frac{{\beta k \Theta }}{{{{\left( {{\mu _k} + \beta
k } \right)}^2}}},
\end{equation}
and
\begin{equation}\label{eq19}\tag{S19}
\begin{split}
\frac{{\partial \mathcal {L}}}{{\partial \Theta }} =  &\sum\limits_k
{\frac{{\beta kP(k)}}{{{\mu _k} + \beta k\Theta }} - } \sum\limits_k
{\frac{{{\beta ^2}{k^2}P(k)\Theta }}{{{{\left( {{\mu _k} + \beta
k\Theta } \right)}^2}}}}  \\& - \kappa \sum\limits_k
{\frac{{kP(k)}}{{\left\langle k \right\rangle }}\frac{{{\beta
^2}{k^2} \Theta }}{{{{\left( {{\mu _k} + \beta k\Theta }
\right)}^2}}}}.
\end{split}
\end{equation}
Taking the derivative of $\mathcal {L}$ with respect to the Lagrange
multipliers $\tau$ and $\kappa$, we obtain the constraint equation
Eq.(\ref{e8}) and the self-consistent equation Eq.(\ref{e12}) of
$\Theta$ in the main text.

Letting $\partial \mathcal {L}/\partial \mu_k=0$, we obtain
\begin{equation}\label{eq20}\tag{S20}
\mu _k = \left\{ \begin{gathered}
  \sqrt {\frac{{\beta k\Theta }}{\tau } + \frac{{\kappa \beta {k^2}}}{{\tau \left\langle k \right\rangle }}}  - \beta k\Theta >0,  \; k <k_c \hfill \\
  0, \; k \geq k_c \hfill \\
\end{gathered}  \right.
\end{equation}
where $k_c$ is a threshold degree to guarantee $\mu_k>0$ for $k<k_c$
and it will be determined later. Substituting Eq.(\ref{eq20}) into
Eq.(\ref{eq19}) and letting $\partial \mathcal {L}/\partial
\Theta=0$, we obtain
\begin{equation} \label{eq21}\tag{S21}
\begin{split}
\sqrt {\frac{{\beta \tau \left\langle k \right\rangle }}{\Theta }}
\sum\limits_{k = {k_{\min }}}^{{k_c}} {\frac{{P(k)\sqrt k }}{{\sqrt
{\left\langle k \right\rangle  + \kappa k} }}}  - \beta \tau
\left\langle k \right\rangle \sum\limits_{k = {k_{\min }}}^{{k_c}}
{\frac{{P(k)k}}{{\left\langle k \right\rangle  + \kappa k}}}
 \\- \beta \kappa \tau \sum\limits_{k = {k_{\min
}}}^{{k_c}} {\frac{{P(k){k^2}}}{{\left\langle k \right\rangle
 + \kappa k}}  - \frac{\kappa }{{\left\langle k
\right\rangle \Theta }}\sum\limits_{k = {k_c}}^{{k_{\max }}} {kP(k)}
}  = 0. \nonumber
\end{split}
\end{equation}
Combining Eq.(\ref{e8}) in the main text and Eq.(\ref{eq20}), we
obtain
\begin{equation}\label{eq22}\tag{S22}
\begin{split}
\mu  =  &\sqrt {\frac{{\beta \Theta }}{{\tau \left\langle k
\right\rangle }}} \sum\limits_{k = {k_{\min }}}^{{k_c}} {P(k)\sqrt k
\left( {\sqrt {\left\langle k \right\rangle  + \kappa k} } \right) }
\\&-\beta \Theta \sum\limits_{k = {k_{min}}}^{{k_{c}}} {P(k)k}.
\end{split}
\end{equation}
Combining Eq.(\ref{e12}) in the main text and Eq.(\ref{eq20}), we
obtain
\begin{equation}\label{eq23}\tag{S23}
\Theta= \sqrt {\frac{{\beta \tau \Theta }}{{\left\langle k
\right\rangle }}} \sum\limits_{k = {k_{\min }}}^{{k_c}}
{\frac{{P(k){k^{3/2}}}}{{\sqrt {\left\langle k \right\rangle  +
\kappa k} }}}  + \frac{1}{{\left\langle k \right\rangle
}}\sum\limits_{k = {k_c}}^{{k_{\max }}} {kP(k)  }.
\end{equation}

\end{document}